\documentclass{sigchi}
\pdfoutput=1 

\toappear{Permission to make digital or hard copies of all or part of this work for personal or classroom use is granted without fee provided that copies are not made or distributed for profit or commercial advantage and that copies bear this notice and the full citation on the first page. Copyrights for components of this work owned by others than the author(s) must be honored. Abstracting with credit is permitted. To copy otherwise, or republish, to post on servers or to redistribute to lists, requires prior specific permission and/or a fee. Request permissions from Permissions@acm.org. \\
{\emph{CSCW'17}}, February 25--March 01, 2017, Portland, OR, USA. \\
Copyright is held by the owner/author(s). Publication rights licensed to ACM. \\
ACM 978-1-4503-4335-0/17/03. \$15.00   \\
DOI: http://dx.doi.org/10.1145/2998181.2998196}

\usepackage{balance}  
\usepackage{graphics} 
\usepackage{times}    
\usepackage{url}      
\usepackage{booktabs}
\usepackage{tabulary} 
\usepackage{tablefootnote} 
\usepackage{comment}
\usepackage{quoting}
\usepackage{todonotes}
\usepackage[export]{adjustbox}
\usepackage{enumitem}

\makeatletter
\def\url@leostyle{%
  \@ifundefined{selectfont}{\def\UrlFont{\sf}}{\def\UrlFont{\small\bf\ttfamily}}}
\makeatother
\def\pprw{8.5in}
\def\pprh{11in}

\usepackage[pdftex]{hyperref}
\hypersetup{
pdftitle={Mechanical Novel: Crowdsourcing Complex Work through Reflection and Revision},
pdfauthor={Joy Kim, Sarah Sterman, Allegra Argent Beal Cohen, Michael S. Bernstein},
pdfkeywords={Social computing, online creative collaboration, crowdsourcing, storytelling},
bookmarksnumbered,
pdfstartview={FitH},
colorlinks,
citecolor=black,
filecolor=black,
linkcolor=black,
urlcolor=black,
breaklinks=true,
}

\newcommand\tabhead[1]{\small\textbf{#1}}

\begin{document}

\title{Mechanical Novel: Crowdsourcing Complex Work\\ through Reflection and Revision}

\numberofauthors{1}
\author{
  \alignauthor Joy Kim, Sarah Sterman, Allegra Argent Beal Cohen, Michael S. Bernstein\\
     \affaddr{Stanford University}\\
     \email{\{jojo0808, ssterman, aacohen\}@stanford.edu, msb@cs.stanford.edu} \\
}

\maketitle

\begin{abstract}
Crowdsourcing systems accomplish large tasks with scale and speed by breaking work down into independent parts. However, many types of complex creative work, such as fiction writing, have remained out of reach for crowds because  work is tightly interdependent: changing one part of a story may trigger changes to the overall plot and vice versa. Taking inspiration from how expert authors write, we propose a technique for achieving interdependent complex goals with crowds. With this technique, the crowd loops between reflection, to select a high-level goal, and revision, to decompose that goal into low-level, actionable tasks. We embody this approach in Mechanical Novel, a system that crowdsources short fiction stories on Amazon Mechanical Turk. In a field experiment, Mechanical Novel resulted in higher-quality stories than an iterative crowdsourcing workflow. Our findings suggest that orienting crowd work around high-level goals may enable workers to coordinate their effort to accomplish complex work.

\end{abstract}

\keywords{
	Social computing; online creative collaboration; crowdsourcing; storytelling.
}

\category{H.5.3.}{Group and Organization Interfaces}{Collaborative computing}

\section{Introduction}

\begin{quote}
\textit{I know very dimly when I start what's going to happen. I just have a very general idea, and then the thing develops as I write.
---Aldous Huxley \cite{aldoushuxley}}
\end{quote}


Crowdsourcing platforms such as Amazon Mechanical Turk bring together tens to thousands of people to accomplish complex work at massive scale, allowing the crowd to collaborate on goals such as researching purchases \cite{Kittur:2011:CCC:2047196.2047202}, classification tasks \cite{Simpson:2014:ZOW:2567948.2579215}, and even creating music videos \cite{johnnycash}. 
Currently, crowdsourcing systems accomplish these types of large tasks by decomposing work into independent microtasks. 
These microtask systems present work in an assembly line-like structure called a workflow \cite{bigham2014human}, using mechanisms such as iteration \cite{Little:2010:EIP:1837885.1837907}, clustering \cite{Chilton:2013:CCT:2470654.2466265}, voting \cite{Little:2010:THC:1866029.1866040}, and other patterns for splitting work \cite{Bernstein:2010:SWP:1866029.1866078,Kittur:2011:CCC:2047196.2047202,Kulkarni:2012:CCW:2145204.2145354}. 
Because these microtasks are independent, crowd workers can complete work without worrying about how their contributions affect others. As a result, large goals can be achieved quickly and at scale.

However, effective workflows are difficult to create in advance. To design a workflow of microtasks, an expert must first form a well-defined problem, then engage in an expensive and time-consuming process where they repeatedly test and iterate on potential workflow designs. Furthermore, the expert may run into common problem-solving barriers such as design fixation \cite{jansson1991design}, difficulty decomposing work into microtasks \cite{Kim:2014:EEC:2531602.2531638}, and fear of failure \cite{bayles2012art}.
This process is difficult for crowds as well: systems like CrowdForge \cite{Kittur:2011:CCC:2047196.2047202} and Turkomatic \cite{Kulkarni:2012:CCW:2145204.2145354} have explored how the crowd can dynamically help experts decide how to partition work, but have found that workers require expert intervention \cite{Kulkarni:2012:CCW:2145204.2145354} or a high-level initial decomposition of tasks \cite{Kittur:2011:CCC:2047196.2047202} in order to decompose work without derailing from the intended goal.

\begin{figure*}
\centering
\includegraphics[width=1\textwidth]{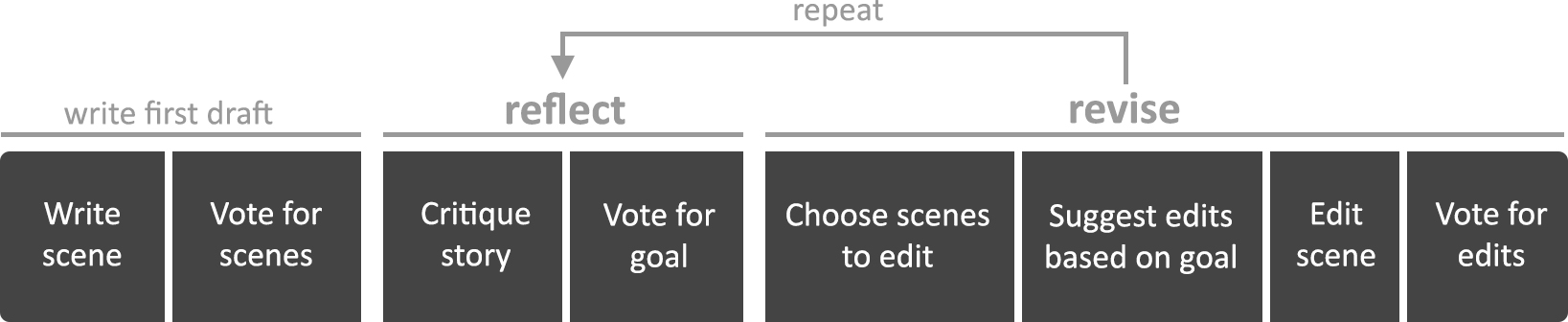}
\caption{Mechanical Novel's crowdsourcing loop alternates between high-level reflection to set a goal, and low-level revision to execute that goal.}
\label{fig:mnovel_flow}
\end{figure*}


In contrast, skilled creators iteratively create and revise goals to develop their vision as they work \cite{flower1981cognitive, sharples1999we}. That is, they know that problems are not always well-defined and that they may need to make many attempts before a solution becomes clear.
With this in mind, we introduce a technique for continually updating and executing high-level goals with crowds.
Rather than asking the crowd to help decompose a static goal, this technique loops between two phases: \emph{reflecting} on the crowd's progress so far to brainstorm and choose a high-level goal, and \emph{revising} the artifact by decomposing that goal into actionable, low-level tasks through which workers make edits.
For example, crowdworkers writing a short story could decide that a story ends too abruptly, and act on that in a specific way by brainstorming a different ending. This new goal can guide workers in deciding how other parts of the story need to change and unlock appropriate parts of the story for editing.
Each goal can still be decomposed into microtasks, making this approach usable in existing crowdsourcing environments.


We instantiate our crowdsourcing strategy of reflection and revision in \textit{Mechanical Novel}, a system that coordinates crowd workers from Amazon Mechanical Turk to write short fiction stories. 
Fiction writing was chosen as a test domain due to the difficulty of defining clear expected solutions (i.e., many different types of stories are acceptable instantiations of an initial idea) and its inherent resistance to being broken down into independent subtasks. For this reason, collaboratively writing high-quality stories has been repeatedly explored by previous work \cite{mason2008million,foldingstory,Kim:2014:EEC:2531602.2531638,Kim:2016:SSS:2818048.2820072} but has remained out of reach for crowds without the help of a leader. 

In Mechanical Novel, after first creating an initial first draft of a story based on a story prompt, workers \emph{select a goal} by reflecting on their progress on the work so far: workers generate critiques, which includes suggesting a possible direction for how the story could change (e.g., foreshadow the death of a love interest).
After voting among these suggestions to choose the next high-level goal to work towards, workers then \emph{execute the goal}. Workers select which parts of the story need to change in order to address the high-level goal, and suggest a specific change for each part of the story they selected (e.g., the love interest says, ``I will always be here for you''). This decomposes the high-level goal into specific tasks the crowd can act on. Workers then vote on these low-level suggestions, and revise portions of the story based on these tasks. The process then repeats, allowing the crowd to further improve the story by selecting a new goal to pursue.

In a controlled study comparing an iterative crowdsourcing workflow with Mechanical Novel, Mechanical Novel produced stronger stories as rated by readers. Specifically, Mechanical Novel's stories had stronger plots (with clearer beginnings, middles, and ends). In iterations on six story drafts with known narrative problems, Mechanical Novel identified and successfully fixed high-level problems with plot and character, in contrast to the iterative workflow's focus on spelling and grammar.

In summary, this paper makes the following contributions:
\begin{itemize}[noitemsep]
\item The \emph{reflect and revise} crowdsourcing technique, which enables crowds to collectively monitor their progress and flexibly contribute work based on high-level goals of their choosing.
\item \emph{Mechanical Novel}, an example system that demonstrates this technique in the context of storywriting, a domain that has typically remained out of reach for crowdsourcing systems.
\item An evaluation of Mechanical Novel that shows the reflect and revise technique can generate short stories with stronger high-level characteristics (such as plot and character) than stories generated by a control system.
\end{itemize}

Crowds that are able to collectively articulate and execute high-level goals as they work could enable not just collaborative fiction-writing but a new class of crowd-powered work, including breaking news stories that are revised in real-time as new information appears, or reworking films across several stages or mediums (e.g., from a script to a storyboard to video).

\section{Related Work}
We focus on developing techniques that allow the crowd to select and act on high-level goals. To inform our design, we examine the strengths and limitations of how crowds work together in existing collaborative environments.

\subsection{Collaborating through context-free tasks}
People often divide collaborative writing work by identifying sections of text that are independent from each other, and then working in parallel on a single document or writing in turn \cite{Kim:2001:RPC:568755.568759,Noel:2004:ESC:967836.967837}.
Many crowdsourcing strategies think about tasks in a similar manner. In these, the role of subtasks is to produce sub-results that are mergable into the final result: crowdworkers caption sections of a speech by captioning one small snippet at a time \cite{Lasecki:2012:RCG:2380116.2380122};
flash teams frame collaborative expert crowd work around sequences of linked tasks and finding appropriate inputs and outputs from one phase to another \cite{Retelny:2014:ECF:2642918.2647409}; workers create a music video by drawing one video frame at a time \cite{johnnycash}; and still other work propose patterns \cite{Bernstein:2010:SWP:1866029.1866078, Kittur:2011:CCC:2047196.2047202, Kulkarni:2012:CCW:2145204.2145354} for breaking down complex tasks into context-free subtasks. 
These workflows can often produce complex work more quickly or more accurately than a person working alone.

Another approach is iterative crowdsourcing, where, rather than stopping after a result is put together piece by piece, one worker creates a first draft of the task, and later workers improve it with subsequent tasks. This is already visible in wiki and open source collaborations, where contributors base their own work on work by others. In tasks such as writing factual descriptions, transcribing blurry text, and brainstorming, iterative crowdsourcing processes can improve the quality of work over time \cite{Little:2010:EIP:1837885.1837907}.

At the same time, these workflows are fragile because they cannot flexibly react to change. Results put together piece-by-piece or in parallel may not be coherent, and iterative processes may fixate on improving low-quality work rather than restarting to find a stronger concept \cite{Little:2010:EIP:1837885.1837907}. Similar problems can be seen in existing collaborative storytelling platforms online, which are often implementations of round-robin storytelling games \cite{foldingstory} that do not allow contributors to alter work that has previously been submitted; a new character introduced on a whim by one contributor unilaterally affects all later contributions whether it is good for the story or not.
In other words, workflows 
lack support for reciprocal interdependence \cite{thompson1967organizations}, where changing one part of the work may necessitate changes to other parts at any time.
Mechanical Novel, instead, supports reciprocal interdependence by allowing workers to revisit and amend the high-level goals toward which they're working.



\subsection{Crowdsourcing with global goals in mind}
To accommodate the unique requirements of complex creative and open-ended work, new crowdsourcing techniques consider global goals (rather than just local ones) by allowing workers to participate in how work is merged. For example, workers can combine the best contributions from multiple past workers \cite{Yu:2011:CCC:1978942.1979147} or repurpose old work for a new goal \cite{Hill:2013:CCC:2441776.2441893}.
Other techniques help workers maintain global consistency: in classification tasks, context regarding the taxonomy developed so far is provided to workers as they arrive to complete tasks in order to allow workers to consider existing categories as they classify items \cite{Chilton:2013:CCT:2470654.2466265, Andre:2014:CSE:2531602.2531653}. Context trees \cite{ilprints1105} recursively merge subparts of a long story to gather an emergent understanding of the larger plot; this strategy explicitly shifts from looking at low-level input to the larger story structure and vice-versa, but does not allow workers to modify the story or summary. Voting on how to keep work consistent and organizing high-level ideas prior to work can also help workers think about work from a global standpoint \cite{Hahn:2016:KAB:2858036.2858364}.

Another body of past work focuses on allowing workers to self-coordinate. In these, tasks are generated---either automatically or by a human leader---according to overall requirements and are made available for workers to take. The crowdware paradigm \cite{Zhang:2012:HCT:2207676.2207708} proposes use of a shared 
todo list of collaboration tasks to solve global constraints in tasks that are hard to decompose (such as planning travel). Apparition \cite{Lasecki:2015:ACU:2702123.2702565} 
features a self-coordinating crowd, 
but workers do not directly reflect on their own organizational strategies, nor can they alter the directions laid out by the designer. 
The MicroWriter \cite{Teevan:2016:SCW:2858036.2858108} similarly focuses on scaffolding direct, co-located collaboration between non-crowd groups, providing a shared space to generate, organize, and act on ideas. This shared space allowed pre-existing groups to benefit from a bottom-up approach of building ideas into written paragraphs through microtasks. Mechanical Novel explores a complementary top-down approach where workers first select a goal based on previous work in order to minimize the effort required to coordinate an unaffiliated crowd.

In other work, leaders and collaborators work together more directly; in animation production \cite{Luther:2013:RLO:2441776.2441891}, writing \cite{Kim:2014:EEC:2531602.2531638, Nebeling:2016:WCW:2858036.2858169}, and ideation \cite{chan2014ideagens}, leaders distribute responsibility by generating tasks around which collaborators focus their efforts. 
However, in these systems, individual changes are requested and vetted by the same person, and contributors are often able to directly communicate with the leader. Instead, Mechanical Novel looks at how crowd workers can iteratively collaborate with each other, and introduces a technique for iterating on a central goal without a central creative authority.

Mechanical Novel expands on past research by exploring how evaluating lower-level work against global goals can help crowd workers generate globally consistent output. In addition, workers choose goals themselves.
Based on this, we hypothesize that allowing crowdworkers to influence both high-level and low-level work may help workers converge on a common creative direction. By allowing them to revise, we open opportunities for workers to challenge and change the constraints of their work when appropriate.

\section{Mechanical Novel}
To enable crowds to manage high-level interdependencies as they collaborate on complex work, we introduce a crowdsourcing technique consisting of two phases. First, crowd workers \emph{reflect} to brainstorm and choose a high-level goal to pursue. Second, workers \emph{revise} their work to achieve this goal by decomposing that goal into specific tasks. This process loops to continually improve previous work.
We test our technique for crowd reflection and revision in a system for collaborative fiction writing called Mechanical Novel. In this section, we describe the workflow (Figure \ref{fig:mnovel_flow}) that guides the crowd through a collaborative revision process. 

\subsection{Designing workflows based on expert practice}
Our technique takes inspiration from expert creative practice.
Experts do indeed break down their work into smaller parts, but not independent tasks: rather, they continuously reflect on their work and use that reflection to revise their goals and decide what to do next \cite{flower1981cognitive}.
An author, for example, does not finish a story after simply linearly filling in a plot outline---they instead write and rewrite while continually reflecting on what their vision is and how to achieve it~\cite{sharples1999we}.
Similar processes occur across many creative domains such as art, architecture, and writing~\cite{schon1983reflective,flower1981cognitive,alexander1977pattern}.
This process, looping between \emph{reflecting} on progress to identify a goal and \emph{revising} based on that goal, allows experts to ``converse'' with their work \cite{schon1983reflective} and evaluate options by trying them out \cite{pecher2005grounding}.
However, because crowd workers are typically not domain experts, this strategy needs to take the form of microtasks in order to use it in crowdsourcing systems. 
Our intent here is not to reduce storytelling to an impassive and mechanic series of steps; in fact, we chose storytelling as an example domain to help us develop the technique we describe in this paper precisely because it requires a flexible process that can respond to flashes of inspiration and emotional sensibility. Designing for non-traditional work tasks (such as writing stories) may uncover new types of crowdsourcing and collaboration techniques that preserve the ability to respond to creative insight.

\begin{figure}[tb]
\centering
\includegraphics[width=\columnwidth,cfbox=lightgray 0.5pt 0pt]{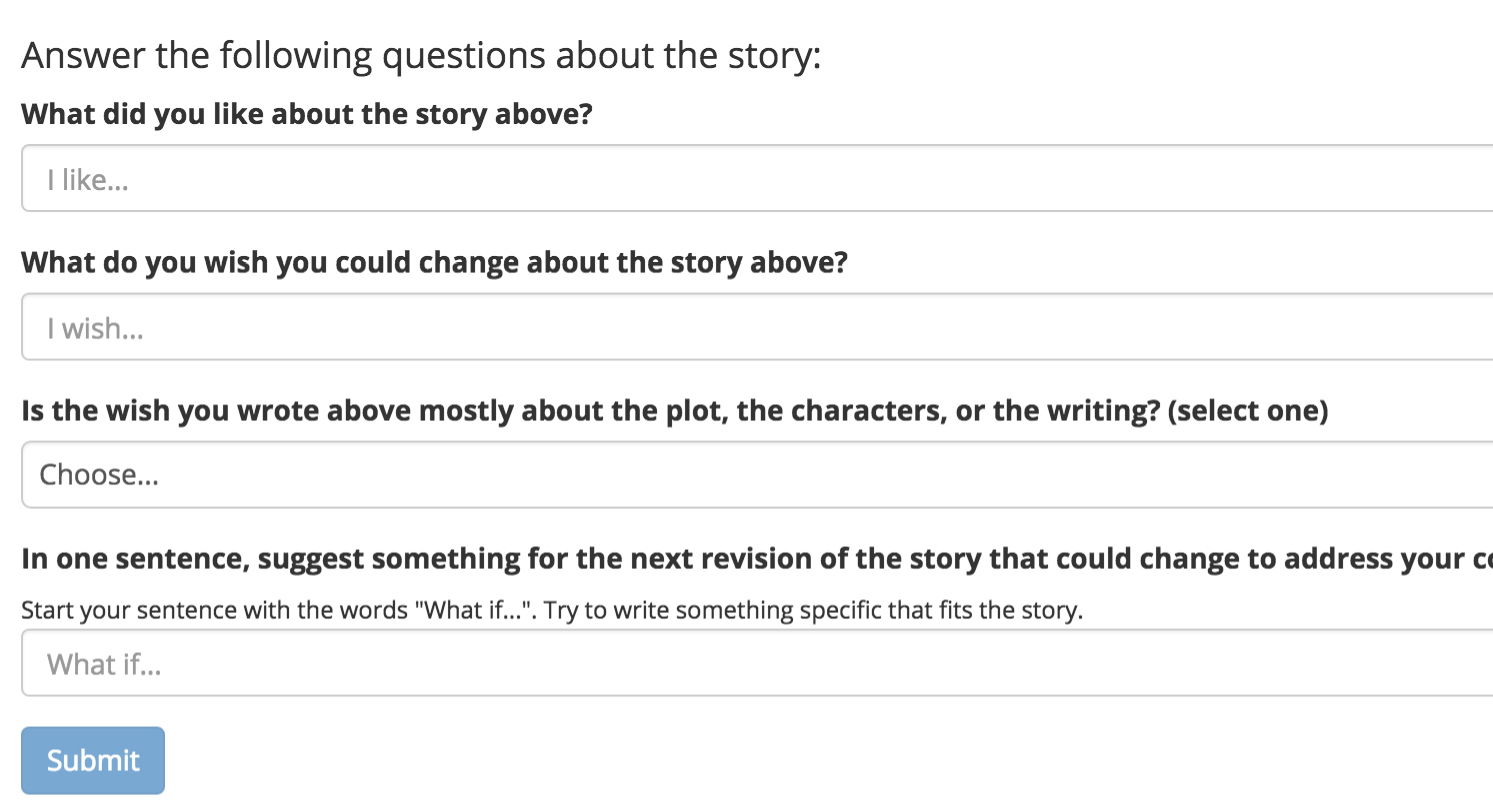}
\caption{Workers critique a story, reflecting on what is working and not working in order to choose a goal for their work.}
\label{fig:mnovel_critique}
\end{figure}

\begin{figure}[tb]
\centering
\includegraphics[width=\columnwidth,cfbox=lightgray 0.5pt 0pt]{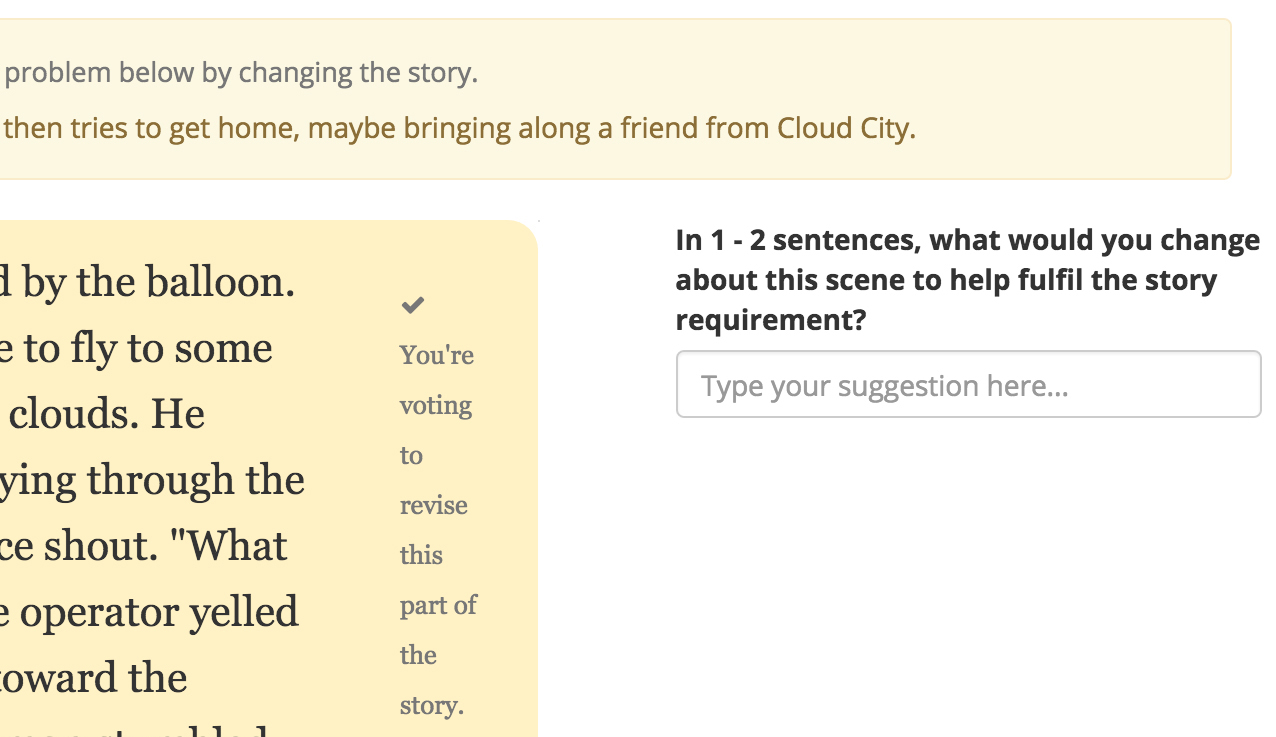}
\caption{Workers select scenes to unlock for revision, suggesting how each scene should change to help achieve the goal.}
\label{fig:mnovel_unlock}
\end{figure}

\begin{figure}[tb]
\centering
\includegraphics[width=\columnwidth,cfbox=lightgray 0.5pt 2pt]{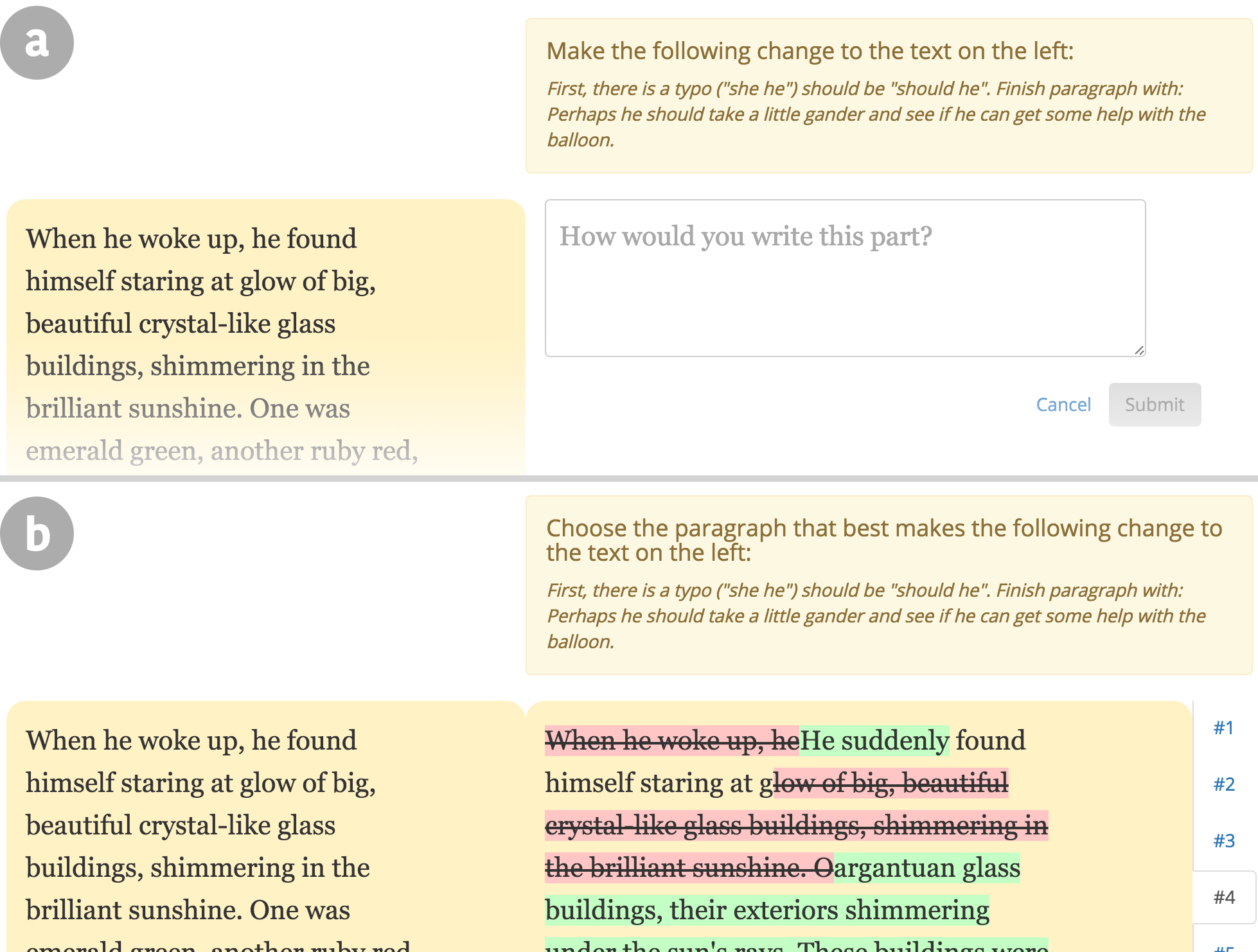}
\caption{Workers (a) propose and (b) vote on candidates for changes to the story based on the high-level goal.}
\label{fig:mnovel_revision}
\end{figure}

\subsection{Initialization: creating a first draft}
For Mechanical Novel to engage in reflection and revision, it must begin with a first draft. The first draft is authored using traditional iterative crowdsourcing strategies. Mechanical Novel initially takes a short prompt describing the overall concept of the story as input, such as ``A young boy named Malcolm finds himself alone in a runaway hot air balloon and accidentally travels to a city in the sky.'' Based on this prompt, the crowd generates the first draft of a story that is six \emph{scenes} long.\footnote{
This story length struck a balance between being long enough to make it difficult to coordinate work and short enough to complete in a reasonable amount of time on Mechanical Turk.} Scenes are the basic unit of writing work in Mechanical Novel; rather than allowing workers to edit any part of the text they like, the system restricts workers to editing within one scene during any task. 

To do this, five workers each independently write a candidate for the text of a scene. Other workers then vote for the best candidate, and Mechanical Novel advances to the next scene. Scenes are written sequentially---from the first to the last---rather than in parallel, to aid workers in coordinating lower-level details such as character names, mood, or writing style. Though this sometimes results in chaotic stories that rapidly change direction, forcing sequentiality ensures that workers concretely define possible creative directions that later workers can choose from when deciding how to improve the story.

\subsection{Reflect: choosing a high-level goal}
At this point, the crowd has created a first draft of a story, which is likely rife with narrative inconsistencies. 
To set a high-level goal for subsequent work, we break down the task of reflection into two steps. First, to generate possible goals to pick from, a new set of workers reads the current version of the story and then generates five critiques using the \textit{I like -- I wish -- what if} method (Figure~\ref{fig:mnovel_critique}) \cite{dschool}. Using this method, workers each write one sentence about what they liked about the story (``I like\ldots''), one sentence about what they wish were different about the story (``I wish\ldots''), and one sentence suggesting a concrete change to the story that would make it better (``What if\ldots?'').


Then, to determine which goal is most pressing or interesting to pursue, other workers then vote for the critique they agree with most. The ``what if?'' with the most votes becomes the chosen goal for later work (e.g., ``What if the story ended with Malcolm learning a lesson about the importance of family?''). In this way, workers  identify 
a new goal for work by reacting to the problems present in the current draft.

\subsection{Revise: translate goals into actionable tasks}
The revision phase of work is divided into four steps.
Workers first vote to indicate which of the story's scenes they think must change in order to achieve the goal. Voting for more than one scene indicates that there are dependencies in the story that require multiple parts of the story to change at the same time. For each scene they vote for, workers also must write a short one-sentence suggestion for how that scene must change in order to achieve the goal (e.g., ``Malcolm should apologize to his grandfather in this scene.'') to generate possible revisions to choose from.

Scenes that at least four (out of 10) workers vote for are then unlocked for editing. For each of the unlocked scenes, a new set of workers vote for the suggestion they think best represents how the scene should change. The suggestions with the highest votes for each of the unlocked scenes become tasks that direct how the story should change. 

Mechanical Novel then asks workers to sequentially fix each unlocked scene, presenting to workers both the high-level goal as well as instructions for incorporating the suggestion into the scene. Fixing a scene involves two more tasks similar to those used to write the first draft; multiple workers propose new versions of the scene based on the task's instructions, then other workers vote for the version that best achieves the suggestion and the higher-level goal. This process is repeated across each unlocked scene.
In this way, the high-level goal serves the purpose of restricting the space of possible contributions from workers.

At this point, workers continue to improve the story by returning to the reflection phase, reading the new version of the story and submitting another set of critiques. They then vote for a new high-level goal, split that goal into actionable tasks, and modify the story based on those tasks, resolving different problems with the story with each revision. Currently, story writing stops after a predetermined number of revision rounds, but in future work, Mechanical Novel could allow the crowd to decide when to end the story (e.g., through votes).

\begin{figure}[t]
\centering
\includegraphics[width=\columnwidth,cfbox=lightgray 0.5pt 3pt]{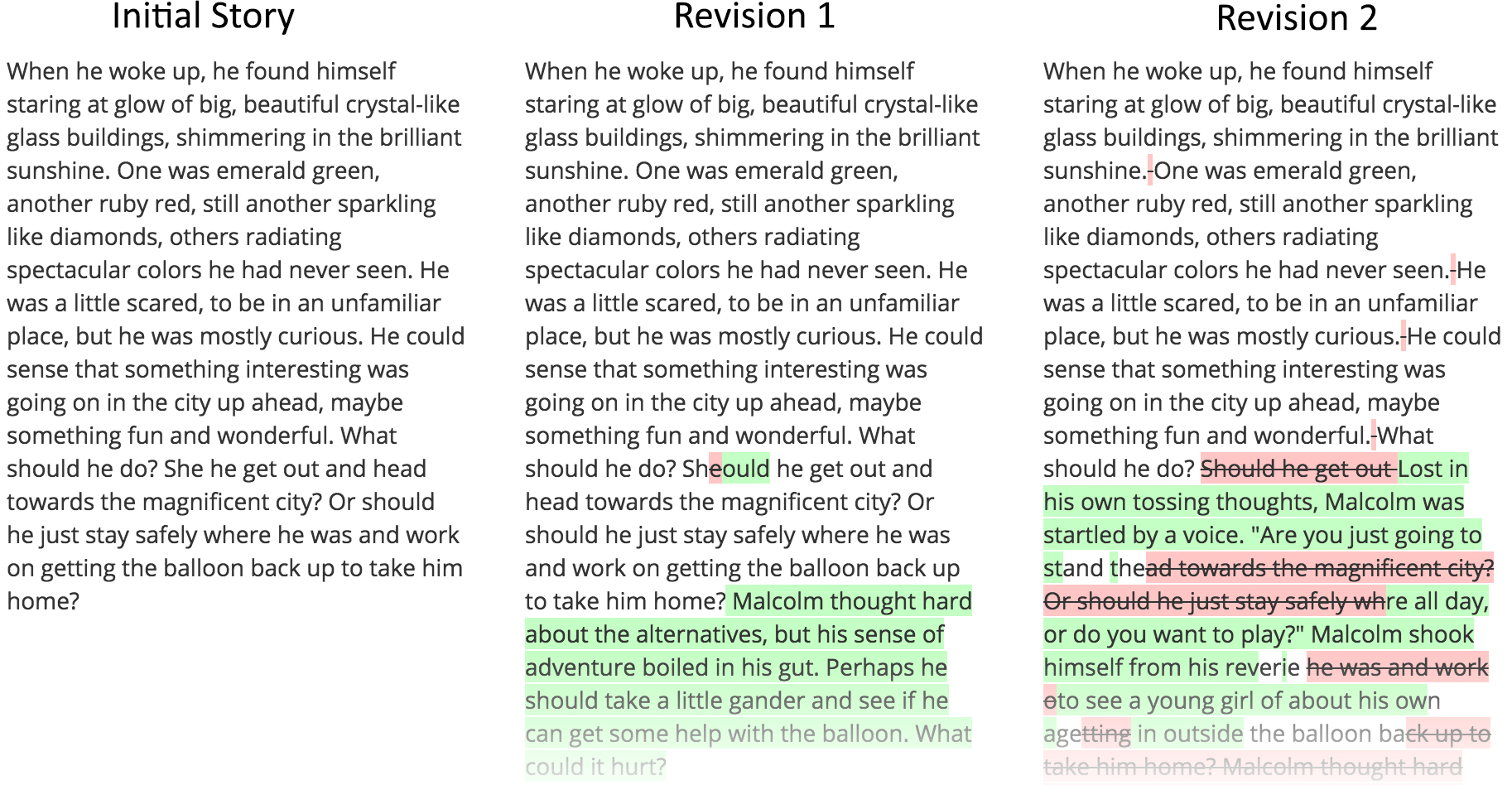}
\caption{A section of a story changing through revisions. Workers first expand this section's ending by having the character make a decision about what to do next, then further expand the story by adding a character who helps progress the story.}
\label{fig:mnovel_changes}
\end{figure}


\section{Evaluation}
Mechanical Novel hypothesizes that structuring work around reflecting and revising high-level goals can allow the crowd to collaborate on complex interdependent work such as fiction writing. In this section, we report on two evaluations exploring whether or not this technique resulted in higher quality stories. In sum, these evaluations find that Mechanical Novel produces stories that were overall preferred over those written using an iterative crowdsourcing strategy, and that it was especially effective at finding and fixing high-level plot issues.

Specifically, the first evaluation gauged how well Mechanical Novel could detect and fix known narrative issues in a series of benchmark stories. The second evaluation compared the quality of stories written by Mechanical Novel and a typical iterative (CrowdForge-style) workflow when given an open-ended story prompt.

\begin{figure*}[t]
\centering
\includegraphics[width=1\textwidth]{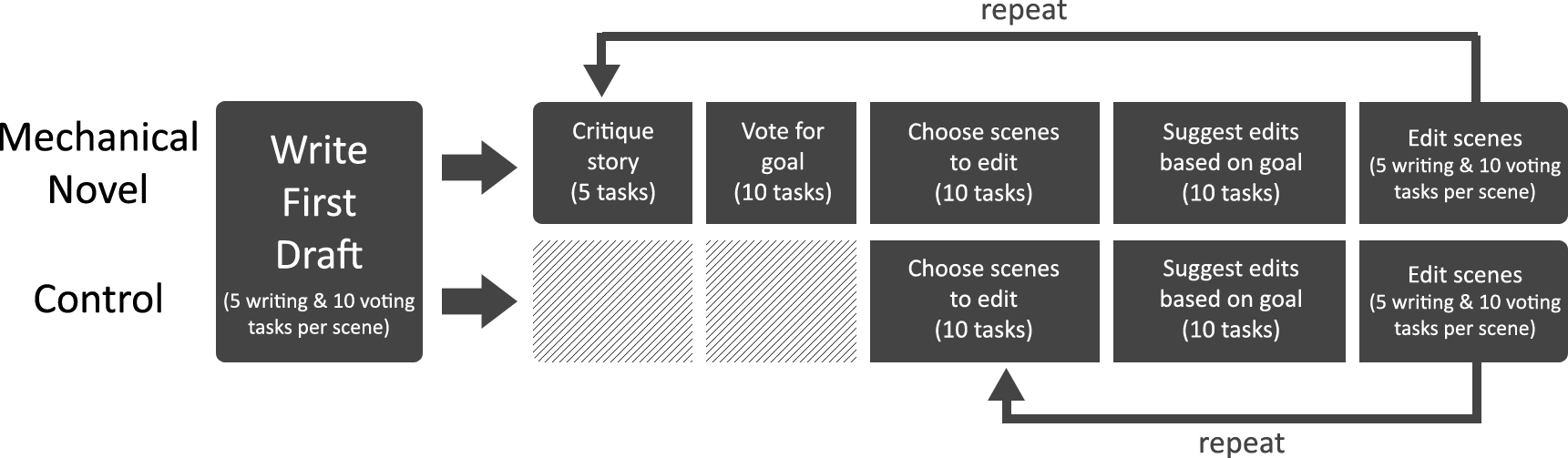}
\caption{The tasks launched on Mechanical Turk for each experimental condition.}
\label{fig:study_conditions_figure}
\end{figure*}

Both studies consisted of two experimental conditions (Figure~\ref{fig:study_conditions_figure}): the \emph{Mechanical Novel} condition, where workers wrote stories by reflecting on a first draft to choose a goal and then revising text, and a \emph{control} condition, where workers wrote stories by voting for which parts of the story to edit and made independent edits to the story's text. The workflows in the Mechanical Novel and control conditions both included tasks where workers unlocked and edited scenes; the workflow for the Mechanical Novel condition included the additional step of reflecting to set a high-level goal.

Figure~\ref{fig:study_conditions_figure} shows the tasks workers did for each revision of stories in each study condition. All tasks were launched simultaneously on Amazon Mechanical Turk to United States workers with a task approval rating of 90\% or higher. Tasks, including those used to generate first drafts, were estimated to take 2 to 8 minutes to complete. Because we wanted to prevent workers from doing tasks from different experimental conditions, we were unable to price different types of tasks individually; instead, we paid all workers based on the longest possible task (priced at \$0.85 each to achieve at an hourly wage of at least the federal minimum wage\footnote{The federal minimum wage at the time of this writing was \$7.25.} on average, in accordance with Mechanical Turk guidelines for academic requesters \cite{dynamo}).
Workers who participated in our tasks were randomly assigned to one of the study conditions for the entirety of their interaction with the system.

\subsection{Benchmark Study}
Our first evaluation sought to measure Mechanical Novel's  performance on six benchmark stories with known narrative issues. This evaluation helps us understand the kinds of high level goals that Mechanical Novel can set and execute. 

We ran the Mechanical Novel and control versions of the system for a single revision cycle over six pre-written benchmark task stories, resulting in 12 stories. Each of the benchmark stories were modified versions of a single short story written by an expert with over 10 years of fiction writing experience (including a crowdfunded, self-published children's novel). Each modified version was changed by the expert to introduce one major problem each (Table \ref{tab:benchmark_table}). We chose both problems that can be fixed independently (such as fixing typos) as well as problems that span across the story (such as changing the way a character speaks) to get a better sense of Mechanical Novel's strengths and weaknesses.

To get a sense of how often workers were able to find problems in the benchmark stories, we tracked the number of times workers correctly voted to change problematic scenes. Two of the authors, blind to condition, then coded edits made to each story in each condition to track how often workers were able to fix the correct problem for each condition ($\kappa = 0.93$). This was repeated three times for each story, so each story had three separate chances to fix errors. Workers were randomized into one of the six stories within each condition, and were only allowed to contribute to one of the repetitions. 

\begin{table}[t!]
  \centering
  \tymin=60pt
  \def\arraystretch{1.3}
  \begin{tabulary}{0.5\textwidth}{LL}
    \toprule
    \tabhead{Problem} &
    \tabhead{Description} \\
    \midrule
    Abrupt ending & The ending of the story is replaced with a sudden exclamation that the story was actually a dream all along. \\
    Extra characters & Dialogue and actions by unnecessary characters are added throughout the story. \\
    Odd dialogue & The main character, who is a child, is changed so that he speaks like an adult. \\
    Point-of-view change & The story changes from third-person to first-person narration halfway through. \\
    Typos & Grammar and spelling errors are introduced to some of the scenes in the story. \\
    Tell, not show & Character's actions are replaced with descriptions of boring or unrealistic behavior (e.g. ``Malcolm's mother held a finger to her lips'' v.s. ``Malcolm's mother told him to be quiet.'') \\
  \bottomrule
  \end{tabulary}
  \caption{The benchmark stories were modified versions of a short story created by an expert, each introducing a common storywriting problem.}
  \label{tab:benchmark_table}
\end{table}

\subsubsection{Results}
The benchmark study suggested that Mechanical Novel is effective at detecting high-level narrative problems (Table \ref{tab:benchmark_find_results_table}). Compared to the control condition, the Mechanical Novel condition resulted in significantly more Turkers correctly voting to change the problematic section of a story for the \textit{abrupt ending} problem ($\chi(1)=4.82, p < 0.05$) and trended towards correctly identifying problematic sections for the \textit{extra characters} problem ($\chi(1)=3.33, p = 0.068$) according to Chi-squared tests. The control condition, on the other hand, identified problematic sections containing lower-level issues such as \textit{typos} ($\chi(1)=10.51, p < 0.01$). Both systems were equally good at detecting \textit{point-of-view changes} ($\chi(1)=1.07, n.s.$) and \textit{odd dialogue} ($\chi(1)=1.40, n.s.$). In general, this reflects the relative strengths of each approach: Mechanical Novel fixed high-level narrative issues, whereas the section-by-section iterative approach fixed low-level technical problems.

Likewise, Mechanical Novel's edits suggested that it can correctly address high-level issues relating to plot and character (Table \ref{tab:benchmark_fix_results_table}), addressing the \textit{abrupt ending} problem 67\% of the time and addressing the \textit{odd dialogue} problem 50\% of the time. However, the low total number of edits makes it difficult to statistically distinguish Mechanical Novel's performance from that of the control workflow, which fixed these problems 25\% and 18\% of the time.

There did not seem to be a difference in how well either system was able to successfully detect or fix the \textit{tell, not show} problem; in addition, while Mechanical Novel correctly identified scenes with the \textit{extra characters} problem, it was not able to correct the issue. This perhaps indicates that, while enabling crowds to think about global elements such as character consistency and plot, Mechanical Novel is less effective at enforcing best practices (such as following the writing rule of ``show, don't tell'') that require workers to be knowledgable and experienced in a domain.

\begin{table}[t!]
  \centering
  \tymin=50pt
  \begin{tabulary}{0.5\textwidth}{LLLLLLL}
    \toprule
    \tabhead{Problem} &
      \multicolumn{3}{c}{\tabhead{Control}} &
      \multicolumn{3}{c}{\tabhead{MNovel}} \\
      & \tabhead{N} & \tabhead{Correct} &
      \tabhead{\%} & \tabhead{N} & \tabhead{Correct} & \tabhead{\%} \\
    \midrule
    Abrupt ending & 30 & 9 & 30\% & 31 & 19 & 61\% \\
    Extra characters & 30 & 13 & 43\% & 30 & 21 & 70\% \\
    Odd dialogue & 31 & 15 & 48\% & 30 & 20 & 67\% \\
    POV change & 30 & 12 & 40\% & 30 & 17 & 57\% \\
    Typos & 30 & 15 & 50\% & 32 & 3 & 9\% \\
    Tell, not show & 31 & 7 & 23\% & 30 & 7 & 23\% \\
  \bottomrule
  \end{tabulary}
  \caption{The total votes cast by workers choosing which sections of the benchmark stories to edit, as well as the number of votes correctly identifying problematic story sections.}
  \label{tab:benchmark_find_results_table}
\end{table}

\begin{table}[t!]
  \centering
  \tymin=50pt
  \begin{tabulary}{0.5\textwidth}{LLLLLLL}
    \toprule
    \tabhead{Problem} &
      \multicolumn{3}{c}{\tabhead{Control}} &
      \multicolumn{3}{c}{\tabhead{MNovel}} \\
      & \tabhead{N} & \tabhead{Correct} &
      \tabhead{\%} & \tabhead{N} & \tabhead{Correct} & \tabhead{\%} \\
    \midrule
    Abrupt ending & 8 & 2 & 25\% & 3 & 2 & 67\% \\
    Extra characters & 7 & 2 & 29\% & 4 & 0 & 0\% \\
    Odd dialogue & 11 & 2 & 18\% & 10 & 5 & 50\% \\
    POV change & 5 & 3 & 60\% & 4 & 2 & 50\% \\
    Typos & 6 & 2 & 33\% & 7 & 2 & 29\% \\
    Tell, not show & 7 & 0 & 0\% & 4 & 1 & 25\% \\
  \bottomrule
  \end{tabulary}
  \caption{The total number of edits made by workers to benchmark stories, as well as the number of paragraphs that correctly corrected problematic story sections.}
  \label{tab:benchmark_fix_results_table}
\end{table}
\begin{table}[t!]
  \centering
  \tymin=80pt
  \def\arraystretch{1.3}
  \begin{tabulary}{0.5\textwidth}{LL}
    \toprule
    \tabhead{Title} &
    \tabhead{Prompt} \\
    \midrule
    The Blue Elephant & Kaley is a girl who spends all her time with an old Blue Elephant doll that was passed down from her grandmother. One day, it disappears. \\
    John Dough & A cutthroat businessman realizes that he's dead and has ended up in heaven, but he has unfinished business... \\ 
    The Hot Air Balloon & A young boy named Malcolm finds himself alone in a runaway hot air balloon and accidentally travels to a city in the sky. \\ 
    The High-Waisted Shorts & Emelia and her high school friends hang out on a normal day, when suddenly, she sees the ghost of a girl wearing beautiful flower-print high-waisted shorts. \\ 
    Number 16 & A serial killer has been monitoring his next victim's movements for months. She is a loner and the perfect target. One day she disappears and nobody notices but him.\\
  \bottomrule
  \end{tabulary}
  \caption{Each study condition included five stories, each based on the prompts above. ``Number 16'' was adapted from Reddit's /r/writingprompts.}
  \label{tab:prompts_table}
\end{table}

\subsection{Story Writing Study}
After establishing that Mechanical Novel allows workers to collaborate to identify high-level goals and to execute them, we wanted to understand how well Mechanical Novel would perform not just in terms of correcting high-level errors but in terms of developing stories from scratch compared to a system representing the state of the art.


In order to ensure we would be able to compare the stories generated by Mechanical Novel and the control system, we seeded each system with the same first draft story text. Crowd workers began by generating five first draft stories---one for each of the five story prompts in Table~\ref{tab:prompts_table}. We then duplicated each first draft to create 10 stories total. Five of these stories were then revised by the crowd using the control system, and five of these stories were revised by the crowd using the Mechanical Novel system.
All stories underwent five rounds of revision. Workers who worked on tasks that generated text for the story were also asked to provide feedback on the task they accomplished, asking specifically about what their goals were in writing their contribution as well as what they thought was difficult about the task. Workers were allowed to contribute to more than one story, but stayed in the same study condition across stories.

To evaluate each story for quality, we asked 215 Mechanical Turk workers who had not participated in any of the story writing tasks to compare a random pair of control and Mechanical Novel stories for one of the story prompts. After being shown each version of the story side-by-side (in random order), workers chose which story they thought was better along several dimensions, such as writing style and presence of story structure (Table \ref{tab:rubric_table}). These dimensions were based on guidelines from a popular book on story writing \cite{burroway2003imaginative}. They also chose which of the two stories they liked better overall.

Lastly, we conducted a grounded theory analysis of how stories changed in each condition by coding the types of changes made in each condition as well as the feedback we received from the crowd workers who worked on writing tasks. Two of the authors, blind to condition, also independently coded each dataset according to emergent themes and resolved conflicts through discussion (paragraph edits: $\kappa = 0.74$; critiques: $\kappa = 0.86$; task feedback: $\kappa = 0.61$).

\begin{table*}[t!]
  \centering
  \tymin=50pt
  \def\arraystretch{1.3}
  \begin{tabulary}{\textwidth}{LLLL}
    \toprule
    \tabhead{Category} &
    \tabhead{Question} &
    \tabhead{Control Votes} &
    \tabhead{MNovel Votes} \\
     \midrule
Imagery & Which story uses better imagery and description? A story with good imagery has description that is memorable and makes it easier to imagine what is happening in the story. & 52 & 162*\\
Coherency & Which story is more coherent? A coherent story has details that are consistent. The story makes sense and doesn't meander or jump around without explanation. & 98 & 115 \\ 
Plot & Which story has a more complete plot? A complete plot has a beginning, middle, and end, with a conflict that arises and is resolved by the end of the story. & 67 & 145*\\
Originality & Which story is more original? An original story has a clear, interesting story premise. & 75 & 136* \\
Style & Which story better uses writing style to enhance the telling of the story? A story with good writing style chooses a voice and tone that makes sense given the story's content and contributes to the telling of the story. & 72 & 143* \\ 
Technical & Which story has less grammar and spelling mistakes? & 130* & 82 \\ 
Overall & Which story did you like better, overall? & 82 & 133* \\ 
  \bottomrule
  \end{tabulary}
  \caption{The questions asked to workers who compared the control and Mechanical Novel stories for each story prompt, as well as the number of workers who voted for the Control story or the MNovel story for each question. ($*=p<0.05$)}
  \label{tab:rubric_table}
\end{table*}

\subsubsection{Results}
Five stories were written for the Mechanical Novel and control conditions, resulting in 10 stories total written by crowdworkers on Mechanical Turk. Stories took an average of 11.38 days ($SD=1.42$) to complete (based on the timestamps of the first and last interactions with the story). Stories were generated through a total of 428 Mechanical Turk tasks completed by an average of 224.5 unique workers per story ($SD=15.63$). 

When rating the final stories overall, workers indicated they liked Mechanical Novel stories better (133 votes for Mechanical Novel v.s. 82 votes for the control workflow; $X^2(1)=12.098, p < 0.01$), according to a Chi-squared test.

\emph{Mechanical Novel stories developed story structure.}
Readers rated Mechanical Novel stories as having significantly more complete plots ($X^2(1)=28.698, p < 0.01$)---that is, readers indicated they viewed Mechanical Novel stories as having more of a complete story arc with a beginning, middle, and end compared to their control version counterparts. Readers also rated Mechanical Novel stories as having significantly more original story premises ($X^2(1)=17.635, p < 0.01$). Considering that Mechanical Novel and control stories for the same story prompt started from the same first drafts, this may indicate that revising stories using high-level goals allowed story ideas to develop in more interesting ways, or that Mechanical Novel stories were more successful at maintaining the story idea established in the first draft.

The Blue Elephant story is an example of how Mechanical Novel was able to generate a more complete story arc. In the first draft of the story, the main character (a young girl) realizes her stuffed elephant is gone, looks all over it, and is finally reunited with it after finding that it has come to life. In the control condition, workers attempted to motivate the main character's actions by establishing that the young girl considers her elephant her best friend. They also add a reason for the elephant's disappearance by having the elephant say he had gone on an adventure.

Mechanical Novel workers, in contrast, revised the story's beginning to include a description of how Kaley received the elephant from her grandmother, which was the same doll her recently deceased mother had when she was a little girl. Workers called back to this backstory in the ending of the story, which reveals that Kaley's love for her grandmother is what brought the Blue Elephant to life, threading a specific theme through the whole story and tying it together.

\emph{Mechanical Novel focused on story over proofreading.}
Readers rated the control stories as having fewer grammar and spelling mistakes ($X^2(1)=10.868, p < 0.01$), indicating that the workers in the control condition seemed to focus more on low-level edits and proofreading.
In contrast, readers rated Mechanical Novel stories as having better use of imagery and description ($X^2(1)=56.542, p < 0.01$) and as having writing styles that better matched each story idea ($X^2(1)=23.447, p < 0.01$).
The final versions of Mechanical Novel stories were also significantly longer than the final versions of the control stories ($t(4.77)=3.65, p < 0.05$), with the Mechanical Novel stories having an average of 1010.6 ($SD=226.15$) words, while the final versions of the control stories were an average of 623.8 ($SD=70.47$) words long.

In sum, Mechanical Novel stories seemed to focus on fleshing out the story itself and how it was told, rather than focusing on local fixes such as missing punctuation or awkward sounding sentences. This is corrobrated by the analysis of types of edits that workers made to each story (Table~\ref{tab:edits_table}). We found that workers in the Mechanical Novel condition made significantly more edits that had to do with expanding on descriptions of characters and how they would act ($X^2(1)=12.49, p < 0.01$), while workers in the control condition trended towards more edits related to fixing grammar and spelling ($X^2(1)=3.202, p = 0.074$) and completely reworded paragraphs significantly more often ($X^2(1)=3.95, p < 0.05$). Table~\ref{tab:critiques_table} also shows that Mechanical Novel workers favored high-level goals that improved high-level flow throughout the story over low-level goals (such as correcting spelling and grammar) and goals that would substantially change the story's concept (such as reordering paragraphs).

An example of this can be seen when comparing the Mechanical Novel and control versions of the John Dough story.
The control story starts out with a straightforward description of the character's surroundings:

\begin{quoting}
\emph{John Dough slowly awoke from a foggy haze. He sat up and immediately felt a searing pain shoot through the left side of his body.}

\emph{``Where am I?'' he wondered out loud.
John did not recognize the room he was in. Everything was white and pristine...white walls, white carpet, white couch and white table, and bright white lights. There was no window, and only a single door at the other end of the room.}

---control condition, \emph{John Dough}
\end{quoting}

The Mechanical Novel story, however, uses first-person voice to create vivid imagery of the main character's thoughts and feelings as they wake up in an unfamiliar place:

\begin{quoting}
\emph{I awoke with a start, sitting up abruptly. There was a searing pain shooting through my body.}

\emph{``Where am I?'' I thought to myself.}

\emph{I didn't recognize my surroundings. Everything was white and pristine; white walls, white carpet, white couch, and white table. No windows, a single door across the room... but somehow the room was intensely bright. Strange.}

---Mechanical Novel condition, \emph{John Dough}
\end{quoting}

\begin{table}[!t]
  \centering
  \begin{tabulary}{0.5\textwidth}{LLLLL}
    \toprule
    \tabhead{Edit Type} &
    \tabhead{Control} & 
    \tabhead{\%} &
    \tabhead{MNovel} &
    \tabhead{\%} \\
    \midrule
    Expand characters & 3 & 6\%* & 14 & 37\%* \\
    Improve flow & 4 & 7\% & 8 & 21\% \\
    Add to plot & 11 & 20\% & 5 & 13\% \\
    Clarify or cut text & 10 & 19\% & 5 & 13\% \\
    Add story background & 2 & 4\% & 2 & 5\%\\
    Change story's tone & 1 & 2\% & 2 & 5\% \\
    Rewrite scene & 10 & 19\%* & 1 & 3\%* \\
    Correct technical issues & 9 & 17\% & 1 & 3\% \\
    Emphasize story's moral & 4 & 7\% & 0 & 0\% \\
  \bottomrule
  \end{tabulary}
  \caption{Workers in the Mechanical Novel condition were especially likely to expand characters, while workers in the control condition were more likely to rewrite a scene from scratch. ($* = p < 0.05$)}
  \label{tab:edits_table}
\end{table}

\begin{table}[t!]
  \centering
  \begin{tabulary}{\columnwidth}{LLL}
    \toprule
    \tabhead{Critique Type} & \tabhead{Suggested} & \tabhead{Chosen} \\
    \midrule
    Add to plot & 74 & 7 \\
    Add story background & 16 & 3 \\
    Expand characters & 37 & 2 \\
    Improve flow & 10 & 2 \\
    Emphasize story's moral & 5 & 1 \\
    Correct technical issues & 18 & 1 \\
    Reorder or shorten story structure & 5 & 1 \\
    Clarify or cut text & 11 & 0 \\
    Redo the story's concept & 7 & 0 \\
    Change story's tone & 3 & 0 \\
  \bottomrule
  \end{tabulary}
  \caption{The types of high-level critiques that workers made before starting a revision cycle in the Mechanical Novel condition, as well as the number of times a critique of each type was chosen as a high-level goal.}
  \label{tab:critiques_table}
\end{table}

\begin{table}[!t]
  \centering
  \tymin=50pt
  \def\arraystretch{1.3}
  \begin{tabulary}{0.5\textwidth}{LLLLL}
    \toprule
    \tabhead{Feedback Type} &
    \tabhead{Control} & 
    \tabhead{\%} &
    \tabhead{MNovel} &
    \tabhead{\%} \\
    \midrule    
    Description of changes & 56 & 21\%* & 103 & 40\%* \\
    Inserted new idea & 27 & 10\% & 41 & 16\% \\
    Refined or corrected text & 99 & 36\%* & 29 & 11\%* \\
    Followed suggested changes & 7 & 3\%* & 24 & 9\%* \\
    Improved story pacing & 27 & 10\% & 23 & 9\% \\
    Continued other workers' work & 10 & 4\% & 15 & 6\% \\
    Confusion or frustration with other workers' work & 13 & 5\% & 11 & 4\% \\
    No change needed & 11 & 4\% & 4 & 2\%\\
    Critiqued overall story & 15 & 6\%* & 3 & 1\%* \\
    Set up opportunities for other workers & 3 & 1\% & 3 & 1\% \\
    Too much work to change & 1 & 0.4\% & 0 & 0\% \\
  \bottomrule
  \end{tabulary}
  \caption{Workers in the Mechanical Novel condition were more likely to follow a high-level goal, whereas workers in the control condition were more likely to correct text or attempt to critique the overall story from within a single paragraph or scene. ($* = p < 0.05$)}
  \label{tab:story_difficulty_table}
\end{table}

\emph{Mechanical Novel allowed workers to coordinate.}
Workers in the Mechanical Novel condition encountered less friction in contributing to the story. After analyzing the comments workers wrote after contributing story text (and revisions to text), we found that Mechanical Novel workers were more likely to explain their work as following the suggested changes (as informed by the high-level goal) created by previous workers ($X^2(1)=9.56, p < 0.01$). Workers in the control condition, on the other hand, trended towards being more likely to try and focus the story's direction by introducing significant plot changes or twists through their local contribution ($X^2(1)=3.56, p = 0.06$) and also included more critiques of the overall story in their feedback ($X^2(1)=6.46, p < 0.05$) to justify the text they had written.

Surprisingly, nearly all accepted changes to Mechanical Novel stories were created by unique workers, with an average of 7.8 accepted changes per Mechanical Novel story ($SD = 1.48$) by an average of 7.4 unique workers ($SD = 1.34$). Revisions to control stories were distributed among workers similarly, with an average of 9.4 accepted changes per control story ($SD = 1.52$) by an average of 8.8 unique workers ($SD = 1.92$). There was no significant difference between study conditions in the number of unique workers whose revisions were accepted ($t(7.2)=1.33, n.s.$).
In addition, out of the Mechanical Novel workers who participated in more than one task, 62.2\% participated in both reflect and revision phases of a story. Out of Mechanical Novel workers who completed at least 3 tasks, 92.2\% did at least 2 different types of tasks and 70.6\% did at least 3 different types of tasks.
In other words, it was not the case that a few skilled workers were dominating story-writing tasks in Mechanical Novel.

\emph{Both conditions struggled with coherency.}
There was no significant difference between the control and Mechanical Novel stories in terms of how coherent they were perceived to be ($X^2(1)=1.357, n.s.$)---that is, all stories were seen as lacking consistency in details (for example, in The Blue Elephant, workers did not resolve whether it was Kaley's mother or grandmother who had passed away). In addition, in their feedback, there was no significant difference between conditions on how often workers expressed frustration with having to struggle against earlier or later parts of the story ($X^2(1)=0.009, n.s.$):

\begin{quoting}
\emph{That the person who wrote the paragraph before mine paid no attention to pacing and didn't seem to know much about hot air balloons. The ``accidentally knocked unconscious'' cliche was a bit annoying...}

---Worker, control condition, \emph{The Hot Air Balloon}
\end{quoting}


\section{Discussion}
Through an analysis of how the crowd wrote stories through Mechanical Novel, we found that techniques for setting high-level goals---inspired by expert writers' process---helped the crowd produce stories with stronger narrative arcs and description compared to stories written using a traditional crowdsourcing workflow.

\subsection{Enabling flexbility and encouraging diversity}
We also found that Mechanical Novel spread work across many unique workers, rather than allowing a few skilled workers to dominate the creative process. This indicates that the reflect and revise technique we use in this paper provides a steady source of fresh perspectives on a complex task where creative exploration is necessary.
The diversity of perspectives that this technique affords may expand the types of work crowdsourcing can support. For example, citizen journalism is recognized for its ability to disseminate news faster and with wider reach than mainstream news organizations. At the same time, much like crowdsourcing, it faces criticisms stemming from its decentralized nature; reports by citizen journalists are difficult to regulate and may not adhere to standards of quality, trustworthiness, objectivity, and ethics. While a professional journalist could help solve these problems, the presence of an expert also negates the value of citizen journalism as an alternative source of timely information. Enabling a crowd of decentralized contributors to revise and reflect together on the news they produce may preserve the ability to quickly propogate information while keeping each others' facts and biases in check through brainstorming and voting for a common high-level goal. In addition, the ability to continually revise and act on new goals may allow crowds to work together in generating stories around events such as natural disasters where centralized information is unavailable.

Reflecting and revising on work is also medium-agnostic and can be implemented as part of a crowdsourcing system regardless of the actual work task at hand. With Mechanical Novel, we found that workers submitted and voted for critiques and edits appropriate for story writing (such as those that focused on plot and character) without having the system specify desired input from the crowd. For this reason, one could imagine that the crowd could use this technique to flexibly support work that moves through different stages of production. Reflections on the script for a crowdsourced film, for example, could lead to revisions of a storyboard or casting choices. Then, the actual task of creating the film could be supported through existing crowdsourcing strategies and interfaces (e.g., \cite{johnnycash}).

However, we also found that Mechanical Novel performed less well than the control system when it came to low-level work (such as correcting grammar and spelling errors). This may mean that reflecting and revising could be used in a complementary way with existing crowdsourcing patterns; for example, find-fix-verify \cite{Bernstein:2010:SWP:1866029.1866078} could be used to refine the stories that Mechanical Novel generates.

\subsection{Going beyond short stories}
At the same time, Mechanical Novel is currently limited by its assumption that the short story being generated is small enough to fit in the working memory of each worker. That is, a worker has to be able to read the whole story and make a critique in order to select a high-level goal for subsequent work. In addition, workers currently must be able to look through the entire story to flag which parts of the story must change in order to achieve the high-level goal. For the purposes of exploring the approach of decomposing crowdsourced creative work based on a goal selected by the crowd, we deliberately limited the length of each story so that it is possible for each worker to familiarize themselves with the story in a short amount of time. 

How might this approach be used to generate a larger work? One strategy may be to apply the Mechanical Novel approach recursively, where workers could collaborate on the high-level structure of a story, then dynamically expand on individual chapters or narrative acts.
Another strategy may be to make use of a working memory space for the crowd as seen in past work \cite{Lasecki:2013:CCC:2501988.2502057,Zhang:2012:HCT:2207676.2207708} to further help direct work by letting future workers know the creative intent of past workers.


\subsection{Designing collaboration around reflection and revision}

Why does the reflect and revise technique work? Too much structure can undesirably limit the work that crowd workers do. An early version of Mechanical Novel allowed crowd workers to set high-level goals by having them brainstorm and vote on an outline, much like Crowdforge~\cite{Kittur:2011:CCC:2047196.2047202}. Our intent here was to allow workers to concentrate on brainstorming the bigger picture without having to worry about the details of how the story would actually be written. However, we found that workers would work within the outline far too strictly (similar to worker behavior seen in other highly-structured crowd systems \cite{Kulkarni:2012:CCW:2145204.2145354}). This made it hard for workers to explore a wide range of possible creative directions inspired by the story outline; they would not change much from the initial outline that was selected. This may be because crowd workers may err on the side of caution when told to make changes that may or may not be correct in order to avoid having their work rejected.

Instead, we had to design a way for workers to concretely explore possible creative directions. At first, we tried asking workers to brainstorm a theme or moral for the story that would ground later work. Though workers were generally able to select a reasonable theme to guide the next revision of a story, they had difficulty translating such an abstract high-level idea into concrete changes. Instead, critiques provided a way for workers to think about high-level changes in terms of what they wanted to story to specifically look like after revision took place.
However, this means that the reflect and revise technique only works to the extent that non-experts can make evaluations. For example, the crowd may be able to select reasonable goals for changing the structure and flow of a research paper, but are less likely to assess a research paper in terms of how it compares to existing literature. Techniques such as scaffolding feedback \cite{Xu:2014:VGS:2531602.2531604} could help support the reflect phase of work in more specialized domains such as science or design.

Lastly, Mechanical Novel was designed around the constraints of Mechanical Turk, which rewards crowd workers for quickness and punishes workers for subpar work quality. A new kind of marketplace---perhaps one that encourages slower, thoughtful work or risky brainstorming---may better support the type of creative work described in this paper.
In future work, experiments that probe into the relative difficulty of reflection compared to revision may help define the optimal incentive scheme such a market should provide. For example, revision may benefit from thoughtful and careful work while reflection may work best when workers are asked to make snap decisions (or vice versa); this, in turn, may require different reward systems (such as rewarding based on quantity versus quality).

\section{Conclusion}
In this paper, we enabled crowds to collaborate on complex creative work through a technique where the crowd \emph{reflects} on their work and translates those reflections into concrete \emph{revisions} of the work. When crowdwork is structured around reflection and revision, workers can identify and execute high-level goals even when work cannot be easily split into independent tasks. 
This approach allowed workers to detect and fix high-level storytelling problems and resulted in higher quality stories than those written using a traditional crowdsourcing workflow.
Reflection and revision's focus on high-level work may be an effective complement to existing crowdsourcing techniques.

Mechanical Novel suggests the possibilities that arise if we start to think of crowdwork not just as a collection of tasks to complete but as a collaborative activity that workers themselves can influence. Wisdom---even that of the crowds---comes not from blindly following orders but from dialogue, reflective practice, and revision.

\section{Acknowledgments}
We would like to thank Mechanical Turk workers and study participants for their time and valuable feedback.  Thanks also to our colleagues who helped test early prototypes of Mechanical Novel. Special thanks to Kylie Jue for her help developing the Mechanical Novel system. This material is based upon work supported by the NSF under Grant No. DGE-114747 and Grant No. IIS-1351131 and by the Hasso Plattner Institute Design Thinking Research Program.

\section{Appendix}
Below is one of the stories that workers wrote using Mechanical Novel.

\small
\subsection{The Blue Elephant}
When Kaley was five, she was given a very special gift by her grandmother, a beautiful blue stuffed elephant. This wasn't just any stuffed elephant---it was a  handmade stuffed elephant created for Kaley's mother when she was just a girl, that she had loved dearly. Her mother had passed away when Kaley was a baby and so Kaley was raised by her grandma. The grandmother did her best, but there was always something missing, which made this elephant extra special because it made Kaley feel like she still had part of her mother with her, even though she knew that was crazy. Kaley not only loved her blue elephant because it belonged to her mother, but the elephant also became her best friend. The elephant was always the guest of honor at her tea parties and always slept by her at night and Kaley always felt safe as long as the elephant was with her. 

Kaley often asked her grandma to tell her stories about her mother. She would sit on her lap and hold her elephant while grams told her lovely things about her mom. She loved those precious moments and wanted to ask grandma for a story later on. She couldn't wait! Holding her special elephant and hearing these moments from her mother's life was comforting to her.

What Kaley didn't know was that the elephant was indeed a very special elephant, special beyond her wildest imagination. Because Kaley's mom had loved the elephant so dearly, a part of her had lived on through the elephant. On the morning of Kaley's sixth birthday she woke as the sun danced across her bed and was excited for her party that day. ``Elephant!'' she exclaimed, ``Today is my birthday and we shall have guests, and cake and presents!'' She turned to hug the blue elephant in excitement, but the elephant was gone. ``How strange.'' she thought to herself as she looked to the side of the bed. No elephant there. She climbed down to look under the bed for the elephant but no elephant there either. She sat back on her heels as she was puzzled at where her elephant could be. ``How could it have just disappeared?'' she thought to herself. Kaley was soon to find out how and just how special her blue elephant really was.

It's not like blue elephants just get up and walk away on their own.... or do they? Kaley's blue elephant wasn't like other blue elephants, that's why she always wrote his name in capital letters in her diary and when she wrote short stories at school. ``Blue Elephant'', just like that. That was his name, after all!  Kaley's elephant was blue with a long trunk. She got it from her grandma when she was young and it has been with her ever since. She got it on her 5th birthday as a gift. Maybe he did just get up and walk away. I wouldn't be a bit surprised! I better go look for him right now!

Kaley started her search for the Blue Elephant. First, she searched her room looking in every nook and cranny. No Blue Elephant. After a very long day of searching and not finding Blue Elephant, Kaley started to cry. Kaley's mother had an idea.... She gave Kaley some peanuts to put out to help catch Blue Elephant. Finally Kaley fell asleep for the night. When she woke up in the morning, Blue Elephant was in her bed with a stash of peanuts.

The girl called for the elephant as loudly as she could. He must have heard her, his big ears make it possible to hear from miles away. If he were trapped or something he would surley be able to send a reply with his giant trunk. She wondered where he could be and wandered down the road calling loudly for him. Every few steps she would sit still and listen for him. Then, she thought she heard a muffled reply and put her ear to the ground. She felt the soft thump of an elephant from a far away distance.

Sure enough, Blue was floating gleefully in the pool spraying water triumphantly from his trunk. Kaley could scarcely belief her eyes, but the glee of her imagination took hold and she yelled in joy,

``Blue! Blue! Is that you?''

At the sound of her voice, the little elephant turned his trunk and blew water all over her. Leaving her soaking wet and giggling at her silly little friend.

``How did this happen?'' asked Kaley. The blue elephant was delighted to answer her question. ``You see, Kaley, it was through your love and adoration that I was able to come to life! If it weren't for you I wouldn't be here. Remember that wish you made the day before since? Well, it came true! The spirit of your grandmother lives on, in me. She wanted nothing but for you to be happy. Because of your love I'm here and will answer anything you ask'' 
Shocked, Kaley took a step back and assessed the situation. ``Well, I suppose this wasn't such a bad wish!'' She thought about what she would ask but really all she wanted was to tell her grandma that she missed her ``I miss you grandma, you were gone too soon...''
`` Your grandmother would be happy to hear that Kaley and please tell your mother that she loved her no matter how things turned out''.
`` Elephant? Are you going to stay?'' 
`` I'm afraid not. Grandma's spirit has given me only a temporary time with you and it's just about to expire''
Just like that the dol started to glow and landed in Kaley's hand. Kaley hugged the doll. A doll that she will forever cherish.

\normalsize

%
%
%
%
%
\balance

\bibliographystyle{acm-sigchi}
\bibliography{sample}

\begin{thebibliography}{10}

\bibitem{foldingstory}
Foldingstory.
\newblock \url{http://foldingstory.com/}.

\bibitem{alexander1977pattern}
Alexander, C., Ishikawa, S., and Silverstein, M.
\newblock Pattern languages.
\newblock {\em Center for Environmental Structure 2\/} (1977).

\bibitem{Andre:2014:CSE:2531602.2531653}
Andr{\'e}, P., Kittur, A., and Dow, S.~P.
\newblock Crowd synthesis: Extracting categories and clusters from complex
  data.
\newblock In {\em Proc. CSCW}, CSCW '14, ACM (New York, NY, USA, 2014),
  989--998.

\bibitem{bayles2012art}
Bayles, D., Orland, T., and Morey, A.
\newblock {\em Art \& fear}.
\newblock Tantor Media, Incorporated, 2012.

\bibitem{Bernstein:2010:SWP:1866029.1866078}
Bernstein, M.~S., Little, G., Miller, R.~C., Hartmann, B., Ackerman, M.~S.,
  Karger, D.~R., Crowell, D., and Panovich, K.
\newblock Soylent: A word processor with a crowd inside.
\newblock In {\em Proc. UIST}, UIST '10, ACM (New York, NY, USA, 2010),
  313--322.

\bibitem{bigham2014human}
Bigham, J.~P., Bernstein, M.~S., and Adar, E.
\newblock Human-computer interaction and collective intelligence.
\newblock In {\em Collective Intelligence Handbook}. MIT Press, 2015.

\bibitem{burroway2003imaginative}
Burroway, J.
\newblock {\em Imaginative writing: The elements of craft}.
\newblock Longman, 2003.

\bibitem{chan2014ideagens}
Chan, J., Dang, S., Kremer, P., Guo, L., and Dow, S.
\newblock Ideagens: A social ideation system for guided crowd brainstorming.
\newblock In {\em Second AAAI Conference on Human Computation and
  Crowdsourcing} (2014).

\bibitem{Chilton:2013:CCT:2470654.2466265}
Chilton, L.~B., Little, G., Edge, D., Weld, D.~S., and Landay, J.~A.
\newblock Cascade: Crowdsourcing taxonomy creation.
\newblock In {\em Proc. CHI}, CHI '13, ACM (New York, NY, USA, 2013),
  1999--2008.

\bibitem{dschool}
d.school, S.
\newblock Design method: I like, i wish, what if.
\newblock
  \url{http://dschool.stanford.edu/wp-content/themes/dschool/method-cards/i-like-i-wish-what-if.pdf}.

\bibitem{flower1981cognitive}
Flower, L., and Hayes, J.~R.
\newblock A cognitive process theory of writing.
\newblock {\em College composition and communication\/} (1981), 365--387.

\bibitem{aldoushuxley}
Fraser, R., and Wickes, G.
\newblock Aldous huxley: The art of fiction no. 24.
\newblock {\em The Paris Review 23\/} (1960).

\bibitem{Hahn:2016:KAB:2858036.2858364}
Hahn, N., Chang, J., Kim, J.~E., and Kittur, A.
\newblock The knowledge accelerator: Big picture thinking in small pieces.
\newblock In {\em Proc. CHI}, CHI '16, ACM (New York, NY, USA, 2016),
  2258--2270.

\bibitem{Hill:2013:CCC:2441776.2441893}
Hill, B.~M., and Monroy-Hern\'{a}ndez, A.
\newblock The cost of collaboration for code and art: Evidence from a remixing
  community.
\newblock In {\em Proc. CSCW}, CSCW '13, ACM (New York, NY, USA, 2013),
  1035--1046.

\bibitem{jansson1991design}
Jansson, D.~G., and Smith, S.~M.
\newblock Design fixation.
\newblock {\em Design studies 12}, 1 (1991), 3--11.

\bibitem{Kim:2001:RPC:568755.568759}
Kim, H.-C.~E., and Eklundh, K.~S.
\newblock Reviewing practices in collaborative writing.
\newblock {\em Comput. Supported Coop. Work 10}, 2 (Jan. 2001), 247--259.

\bibitem{Kim:2014:EEC:2531602.2531638}
Kim, J., Cheng, J., and Bernstein, M.~S.
\newblock Ensemble: Exploring complementary strengths of leaders and crowds in
  creative collaboration.
\newblock In {\em Proc. CSCW}, ACM (New York, NY, USA, 2014), 745--755.

\bibitem{Kim:2016:SSS:2818048.2820072}
Kim, J., and Monroy-Hernandez, A.
\newblock Storia: Summarizing social media content based on narrative theory
  using crowdsourcing.
\newblock In {\em Proc. CSCW}, CSCW '16 (2016), 1018--1027.

\bibitem{Kittur:2011:CCC:2047196.2047202}
Kittur, A., Smus, B., Khamkar, S., and Kraut, R.~E.
\newblock Crowdforge: Crowdsourcing complex work.
\newblock In {\em Proc. UIST}, ACM (New York, NY, USA, 2011), 43--52.

\bibitem{johnnycash}
Koblin, A.
\newblock {The Johnny Cash Project}, 2010.
\newblock \url{http://www.thejohnnycashproject.com/}.

\bibitem{Kulkarni:2012:CCW:2145204.2145354}
Kulkarni, A., Can, M., and Hartmann, B.
\newblock Collaboratively crowdsourcing workflows with turkomatic.
\newblock In {\em Proc. CSCW}, CSCW '12, ACM (New York, NY, USA, 2012),
  1003--1012.

\bibitem{Lasecki:2012:RCG:2380116.2380122}
Lasecki, W., Miller, C., Sadilek, A., Abumoussa, A., Borrello, D., Kushalnagar,
  R., and Bigham, J.
\newblock Real-time captioning by groups of non-experts.
\newblock In {\em Proc. UIST}, UIST '12, ACM (New York, NY, USA, 2012), 23--34.

\bibitem{Lasecki:2015:ACU:2702123.2702565}
Lasecki, W.~S., Kim, J., Rafter, N., Sen, O., Bigham, J.~P., and Bernstein,
  M.~S.
\newblock Apparition: Crowdsourced user interfaces that come to life as you
  sketch them.
\newblock In {\em Proc. CHI}, CHI '15, ACM (New York, NY, USA, 2015),
  1925--1934.

\bibitem{Lasecki:2013:CCC:2501988.2502057}
Lasecki, W.~S., Wesley, R., Nichols, J., Kulkarni, A., Allen, J.~F., and
  Bigham, J.~P.
\newblock Chorus: A crowd-powered conversational assistant.
\newblock In {\em Proc. UIST}, UIST '13, ACM (New York, NY, USA, 2013),
  151--162.

\bibitem{Little:2010:EIP:1837885.1837907}
Little, G., Chilton, L.~B., Goldman, M., and Miller, R.~C.
\newblock Exploring iterative and parallel human computation processes.
\newblock In {\em Proc. HCOMP}, HCOMP '10, ACM (New York, NY, USA, 2010),
  68--76.

\bibitem{Little:2010:THC:1866029.1866040}
Little, G., Chilton, L.~B., Goldman, M., and Miller, R.~C.
\newblock Turkit: Human computation algorithms on mechanical turk.
\newblock In {\em Proc. UIST}, UIST '10, ACM (New York, NY, USA, 2010), 57--66.

\bibitem{Luther:2013:RLO:2441776.2441891}
Luther, K., Fiesler, C., and Bruckman, A.
\newblock Redistributing leadership in online creative collaboration.
\newblock In {\em Proc. CSCW}, CSCW '13, ACM (New York, NY, USA, 2013),
  1007--1022.

\bibitem{mason2008million}
Mason, B., and Thomas, S.
\newblock A million penguins research report.
\newblock {\em Institute of Creative Technologies, De Montfort University,
  Leicester, United Kingdom\/} (2008).

\bibitem{Nebeling:2016:WCW:2858036.2858169}
Nebeling, M., To, A., Guo, A., de~Freitas, A.~A., Teevan, J., Dow, S.~P., and
  Bigham, J.~P.
\newblock Wearwrite: Crowd-assisted writing from smartwatches.
\newblock In {\em Proc. CHI}, CHI '16, ACM (New York, NY, USA, 2016),
  3834--3846.

\bibitem{Noel:2004:ESC:967836.967837}
No\"{e}l, S., and Robert, J.-M.
\newblock Empirical study on collaborative writing: What do co-authors do, use,
  and like?
\newblock {\em Comput. Supported Coop. Work 13}, 1 (Jan. 2004), 63--89.

\bibitem{pecher2005grounding}
Pecher, D., and Zwaan, R.~A.
\newblock {\em Grounding cognition: The role of perception and action in
  memory, language, and thinking}.
\newblock Cambridge University Press, 2005.

\bibitem{Retelny:2014:ECF:2642918.2647409}
Retelny, D., Robaszkiewicz, S., To, A., Lasecki, W.~S., Patel, J., Rahmati, N.,
  Doshi, T., Valentine, M., and Bernstein, M.~S.
\newblock Expert crowdsourcing with flash teams.
\newblock In {\em Proc. UIST}, UIST '14, ACM (New York, NY, USA, 2014), 75--85.

\bibitem{schon1983reflective}
Sch{\"o}n, D.~A.
\newblock The reflective practioner.
\newblock {\em London: Temple Smith\/} (1983).

\bibitem{sharples1999we}
Sharples, M.
\newblock {\em An account of writing as creative design}.
\newblock Psychology Press, 1999.

\bibitem{Simpson:2014:ZOW:2567948.2579215}
Simpson, R., Page, K.~R., and De~Roure, D.
\newblock Zooniverse: Observing the world's largest citizen science platform.
\newblock In {\em Proc. WWW}, WWW '14 Companion (2014), 1049--1054.

\bibitem{Teevan:2016:SCW:2858036.2858108}
Teevan, J., Iqbal, S.~T., and von Veh, C.
\newblock Supporting collaborative writing with microtasks.
\newblock In {\em Proc. CHI}, CHI '16, ACM (New York, NY, USA, 2016),
  2657--2668.

\bibitem{thompson1967organizations}
Thompson, J.~D.
\newblock {\em Organizations in action: Social science bases of administrative
  theory}.
\newblock Transaction publishers, 1967.

\bibitem{ilprints1105}
Verroios, V., and Bernstein, M.~S.
\newblock Context trees: Crowdsourcing global understanding from local views.
\newblock In {\em HCOMP 2014} (November 2014).

\bibitem{dynamo}
WeAreDynamo.
\newblock Guidelines for academic requesters.
\newblock
  \url{http://wiki.wearedynamo.org/index.php/Guidelines_for_Academic_Requesters}.
\newblock [Online; accessed 12-May-2015].

\bibitem{Xu:2014:VGS:2531602.2531604}
Xu, A., Huang, S.-W., and Bailey, B.
\newblock Voyant: Generating structured feedback on visual designs using a
  crowd of non-experts.
\newblock In {\em Proc. CSCW}, ACM (New York, NY, USA, 2014), 1433--1444.

\bibitem{Yu:2011:CCC:1978942.1979147}
Yu, L., and Nickerson, J.~V.
\newblock Cooks or cobblers?: Crowd creativity through combination.
\newblock In {\em Proc. CHI}, ACM (New York, NY, USA, 2011), 1393--1402.

\bibitem{Zhang:2012:HCT:2207676.2207708}
Zhang, H., Law, E., Miller, R., Gajos, K., Parkes, D., and Horvitz, E.
\newblock Human computation tasks with global constraints.
\newblock In {\em Proc. CHI}, CHI '12, ACM (New York, NY, USA, 2012), 217--226.

\end{thebibliography}
\end{document}